\documentclass[iop]{emulateapj}
\usepackage{graphicx}
\usepackage{epstopdf}
\usepackage{amsmath, amsthm, amssymb}
\usepackage{multirow}
\usepackage{bbding}
\usepackage{todonotes}

\def\na{NewA}
\newcommand\pc{{\, \rm pc}}
\newcommand\myr{{\, \rm Myr}}

\newcommand\msun{{\, \rm M_\odot}}

\shorttitle{The influence of gas rings on the dynamics of a stellar disk}
\shortauthors{Trani et al.}

\begin{document}


\title{The influence of dense gas rings on the dynamics of a stellar disk in the Galactic Center}


\author{Alessandro A. Trani\altaffilmark{1,2}, Michela Mapelli\altaffilmark{2}, Alessandro Bressan\altaffilmark{1,2}, Federico I. Pelupessy\altaffilmark{3,4}, Arjen van Elteren\altaffilmark{3}, Simon Portegies Zwart\altaffilmark{3}}

\email{aatrani@gmail.com}

\altaffiltext{1}{Scuola Internazionale Superiore di Studi Avanzati (SISSA), Via Bonomea 265, I--34136, Trieste, Italy}
\altaffiltext{2}{INAF-Osservatorio Astronomico di Padova, Vicolo dell'Osservatorio 5, I--35122, Padova, Italy}
\altaffiltext{3}{Leiden Observatory, Leiden University, PO Box 9513, NL-2300 RA, Leiden, The Netherlands}
\altaffiltext{4}{Institute for Marine and Atmospheric Research Utrecht, Department of Physics and Astronomy Utrecht, University Princetonplein 5, 3584 CC Utrecht, The Netherlands}


\begin{abstract}
The Galactic Center hosts several hundred early-type stars, about 20\% of which lie in the so-called clockwise disk, while the remaining 80\% do not belong to any disks. 
The circumnuclear ring (CNR), a ring of molecular gas that orbits the supermassive black hole (SMBH) with a radius of $\sim 1.5 \pc$, has been claimed to induce precession and Kozai-Lidov oscillations onto the orbits of stars in the innermost parsec. 
We investigate the perturbations exerted by a gas ring on a nearly-Keplerian stellar disk orbiting a SMBH by means of combined direct N-body and smoothed particle hydrodynamics simulations.
We simulate the formation of gas rings through the infall and disruption of a molecular gas cloud, 
adopting different inclinations between the infalling gas cloud and the stellar disk. 
We find that a CNR-like ring { is not efficient} in affecting the stellar disk on a timescale of $3 \myr$. In contrast, a gas ring in the innermost $0.5 \pc$ induces precession of the longitude of the ascending node $\Omega$, significantly affecting the stellar disk inclination.
Furthermore, the combined effect of two-body relaxation and $\Omega$-precession drives the stellar disk dismembering, displacing the stars from the disk. 
The impact of precession on the star orbits is stronger when the stellar disk and the inner gas ring are nearly coplanar. We speculate that the warm gas in the inner cavity might have played a major role in the evolution of the clockwise disk.
\end{abstract}





\section{Introduction}

Several hundred young early-type stars lie in the vicinity of SgrA$^\ast$, the compact radio source associated with the supermassive black hole (SMBH) at the center of our Galaxy \citep{kra91,mor93,kra95,gen03,ghe03,sch03,eis05,ghe05,pau06,gil09}. Some of these stars lie in a near-Keplerian disk, called clockwise (CW) disk from the motion that { it shows when projected on} the plane of the sky \citep{pau06,lu09,bar09,do13,lu13,yel14}. 
\citet{bar09} suggest that the CW disk extends up to $0.5 \pc$ and  is  warped or tilted, since the orientation of its normal axis changes by $60^\circ$ from the inner edge to the outer edge. Furthermore, \citet{bar09} suggest the presence of a second dissolving counterclockwise disk. 
However, recent results from \citet{yel14} show that the CW disk extends only up to $0.13 \pc$ and it is neither significantly warped nor tilted. Moreover, \citet{yel14} find no evidence of counterclockwise motion and estimate that only $\sim 20\%$ of the O and Wolf-Rayet stars lie in the CW disk.

The origin of the young stars is puzzling: the tidal shear from the SMBH should disrupt nearby molecular clouds, preventing star formation. However, a disrupted molecular cloud might settle into a disk around the SMBH, which could fragment and form stars (\citealt{bom08,map08,hob09,ali11,map12,map13,luc13,ali13,jal14,map15b}; see \citealt{mapgua15} for a recent review). This scenario can reproduce the observational features of the CW disk, but cannot explain the young stars that do not lie in the disk. A possible scenario to explain such stars is to invoke precession effects, that perturb and dismember the CW disk. 


The SMBH and the young stars are embedded into the cusp of the nuclear star cluster, which has a nearly spherical density profile \citep{gen03,sch07} and is a source of mass precession and two-body relaxation \citep{loc09}. The circumnuclear ring (CNR), a ring of molecular gas that orbits the SMBH with a radius of $\sim 1.5 \pc$, might also perturb the CW disk. In particular, the nearly axisymmetric potential of the CNR might induce precession of the mean orbital elements of a stellar orbit, specifically inclination $i$, eccentricity $e$, argument of periapsis $\omega$ and longitude of the ascending node $\Omega$ \citep{sub09,ulubay09,haa11a,haa11b,sub12,map13}. If the  plane of the orbit and the plane of the axisymmetric potential are misaligned, periodic oscillations in inclination and eccentricity appear. These oscillations, called Kozai-Lidov cycles \citep{koz62,lid62}, might be suppressed by the presence of a strong spherical potential, such as the stellar cusp \citep{iva05,cha09,loc09,sub09}. While these dynamical processes have been investigated for a long time, there is still debate on the actual importance of Kozai-Lidov and precession effects. In particular, the interplay between the CNR and the stellar cusp is still unclear, and previous studies neglected the influence of gas lying in the inner cavity (i.e. inside the CNR). 



Our aim is to investigate the precession induced by dense gaseous rings on the CW disk. We generate such rings self-consistently, by simulating the disruption of a turbulence-supported molecular cloud. In this way, we obtain rings that match the main properties of the CNR, and we also account for the presence of warm gas in the inner cavity \citep{map15}. We investigate precession effects by means of a direct-summation N-body code, coupled with a smoothed particle hydrodynamics (SPH) code thanks to the AMUSE software environment \citep{spz09,spz13,pel13}. In particular, we study the dependence of precession upon the angle between the CW disk and a gas ring. We show that the warm gas that lies in the inner cavity might substantially affect the evolution of the young stars in the Galactic center (GC). In Section~\ref{sec:methods}, we describe the methodology we employed for our simulations; in Section~\ref{sec:results} we present our results. In Section~\ref{sec:discussion}, we discuss the implications of our work and we compare it with previous results. Our conclusions are presented in Section~\ref{sec:conclusions}.

\section{Methods}\label{sec:methods}

\begin{deluxetable}{cccccc}
\tabletypesize{\scriptsize}
\tablecaption{Main properties of the simulations.\label{tab:ic}}
\tablewidth{\linewidth}
\tablehead{
\colhead{Run} & \colhead{Perturber} & \colhead{$\theta_i$} & \colhead{$\theta_{\rm inner}$} & \colhead{Notes}
}
\startdata
A & Yes & 10 & 20 & --\\ 
B & Yes & 45 & 37 & --\\
C & Yes & 90 & 77 & --\\
D & No & -- & -- & --\\ 
A0 & Yes & 10 & 20 & No outer ring\\
A1 & Yes & 10 & 20 & Massless stars\\
\enddata

\tablecomments{\footnotesize 
Column~1: run name; column~2: presence of a perturber (i.e. a molecular cloud falling towards the SMBH); column~3: initial inclination $\theta_i$ between the molecular cloud and the stellar disk; column~4: average inclination $\theta_{\rm inner}$ between the stellar disk and the inner ($r\sim 0.2-0.4 \pc$) gas ring.}
\end{deluxetable}

We make use of the AMUSE software environment \citep{spz09,spz13,pel13} to combine different gas and stellar physics in a single simulation.
AMUSE is a Python framework that embeds several  codes, which are specialized solvers in a single physics domain -- stellar evolution, gravitational dynamics, hydrodynamics and radiative transfer. One of the main features of AMUSE is the ability to couple and run different codes in a single simulation. In particular, it enables the gravitational coupling between the particles of different codes through BRIDGE \citep{fuj07}, which is an extension of the mixed-variable symplectic scheme from \citet{wis91}.

We use the N-body smoothed particle hydrodynamics (SPH) code GASOLINE \citep{wad04,rea10} to simulate the formation of a CNR-like gas ring through the infall and disruption of a molecular gas cloud. The molecular cloud is simulated as in run~{\sc r1} of \citet{map15}, who investigate the formation of circumnuclear disks.  In particular, the molecular cloud is modeled as a spherical cloud with a radius of $15 \pc$ and a total mass of $1.3 \times 10^5 \msun$. Each gas particle has a mass of $1.2 \msun$. 
It has an impact parameter of $b = 26 \pc$ and an initial velocity of $v_{\rm in} = 0.208 v_{\rm esc}$, where $v_{\rm esc}$ is the escape velocity from the SMBH at $25 \pc$. The cloud is seeded with supersonic turbulent velocities and marginally self-bound \citep[see][]{hay11}.
We include radiative cooling, using the same prescriptions as used in \citet{map12}.

GASOLINE uses a kick-drift-kick scheme to integrate the evolution of particles. This scheme is second order accurate in positions and velocities. To achieve higher accuracy in integrating stellar orbits, we calculate the stellar dynamics of a thin stellar disk using the fourth order Hermite N-body code PhiGRAPE and couple this with the time-evolving potential generated by the snapshots of the SPH simulation using a fourth order BRIDGE scheme \citep[][in preparation]{fer15}. The assumption here is that the evolution of the thin stellar disk does not affect the evolution of the gas disk, which is justified by the low mass of the stellar disk compared with the other mass components ($\approx 3 \%$).

To achieve higher accuracy in integrating stellar orbits, we calculate the stellar dynamics of a thin stellar disk using the direct N-body code PhiGRAPE \citep{har07} and couple this with the time-evolving potential generated by the snapshots of the SPH simulation using a fourth order BRIDGE scheme \citep[][in preparation]{fer15}. The assumption here is that the evolution of the thin stellar disk does not affect the evolution of the gas disk, which is justified by the low mass of the stellar disk compared with the other mass components ($\approx 3 \%$).
\begin{figure}[htbp]
  \centering
  \includegraphics[width=\linewidth]{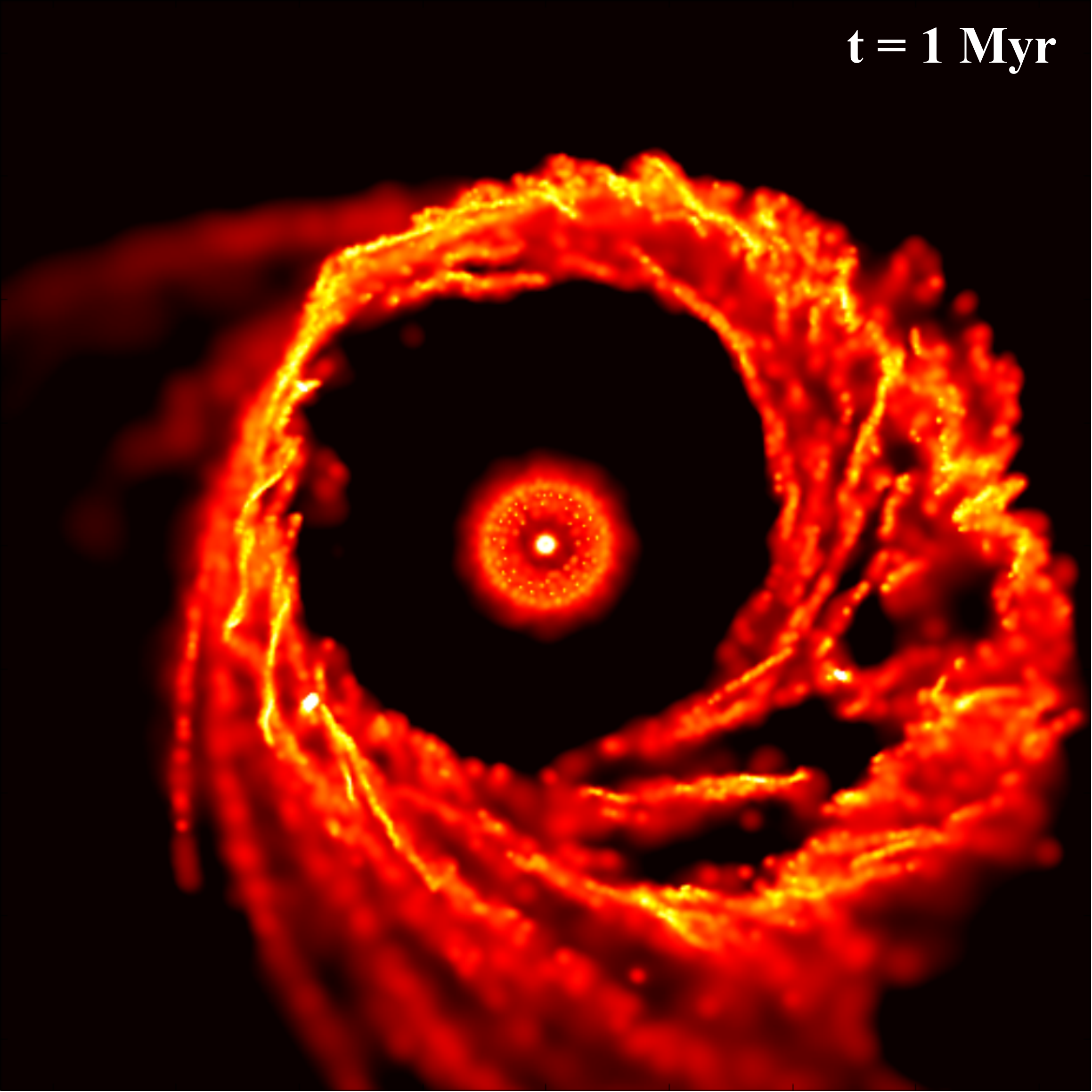}
  \includegraphics[width=\linewidth]{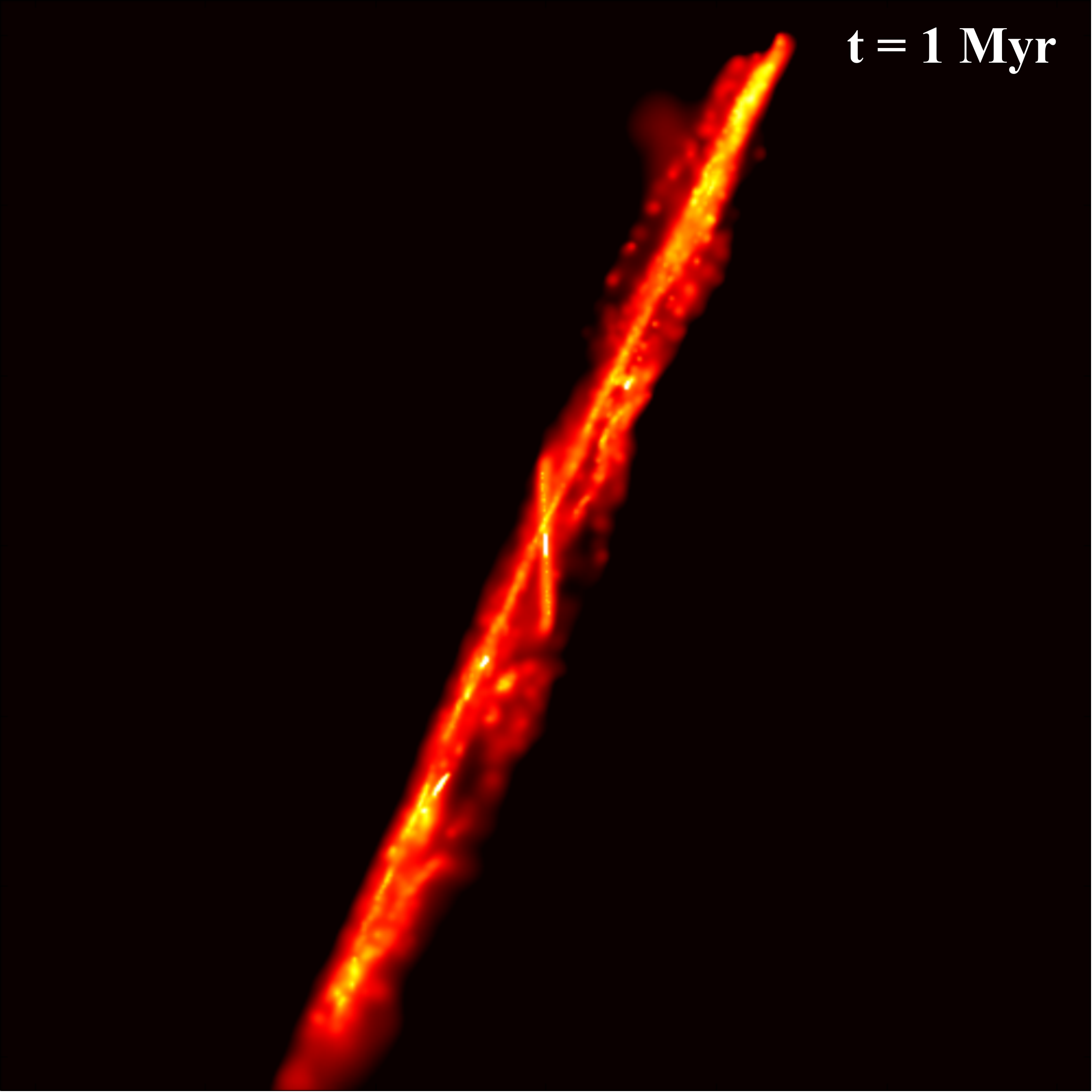}
  \includegraphics[width=\linewidth]{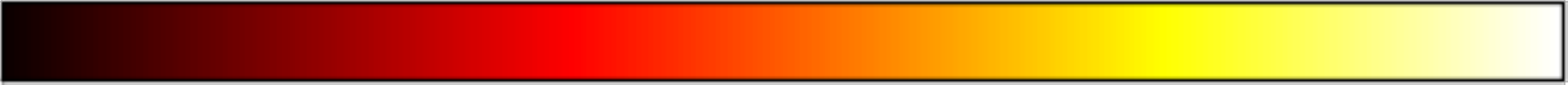}
  \caption{ Color-coded maps of projected density of gas at $1 \myr$ in run~A. Top panel: gas rings seen face-on. Bottom panel: gas rings seen edge-on. The inner gas ring is inclined by $\sim 24^\circ$ with respect to the outer one. The top (bottom) panel measures $8 \pc$ ($6 \pc$) per edge. The color bar ranges from $0.04$ to $49 \msun \pc^{-2}$ (from $0.4$ to $119 \msun \pc^{-2}$) }
\label{fig:gasmap}
\end{figure}
PhiGRAPE uses a fourth order Hermite predictor-corrector scheme to integrate the evolution of the stars. We run PhiGRAPE on GPUs through the SAPPORO library \citep{gab09}. We modified both PhiGRAPE and GASOLINE to include the potential of the SMBH and of the stellar cusp. The SMBH is modeled as a point-mass potential with a mass of $M_{\rm SMBH} = 3.5 \times 10^6 \msun$ \citep{ghe03}. The stellar cusp is modeled as a rigid potential given by a spherical distribution of mass, whose mass density follows the broken-law profile of \citet{sch07}:
\begin{equation}
\rho(r) = 2.8 \times 10^6 \msun \pc^{-3} \left( \frac{r}{r_0} \right)^{-\gamma}
\end{equation}
with $r_0 = 0.22\pc$, $\gamma = 1.75$ for $r>r_0$ and $\gamma = 1.2$ for $r < r_0$, respectively.

For the disk of stars we adopt the outcome of run~E\footnote{ This procedure is not completely self-consistent, because we take the initial conditions for the stellar disk from another simulation, instead of forming the young stars and the gas rings in the same simulation. However, integrating the formation of the stars and studying the dynamical influence of gas rings  in the same simulation is computationally prohibitive. In fact, run~E of \citet{map12} has a factor of 30 higher resolution with respect to the simulations that will be discussed in this paper. Such high resolution is necessary to follow the first stages of the fragmentation process with sufficient accuracy, but run~E of \citet{map12} stalls after $\sim{}5\times{}10^5$ yr. In contrast, to investigate the dynamical effects of gas onto the stellar orbits, we can adopt a lower resolution for the gas component, while we need a much higher accuracy in the integration of stellar dynamics (therefore, we use PhiGRAPE).} of \citet{map12}. In particular, the stellar disk we simulate is composed of 1252 stars with an initial mass function given by a power-law with index $\alpha = 1.5$ and a lower mass limit of $1.3 \msun$. { The total mass of the disk is $4.3 \times 10^3 \msun$, lower than suggested by the most recent observations of the young stars in the central parsec ($\sim{}5-20\times{}10^3 M_\odot$, according to \citealt{lu13}). However, we prefer to use a set of stellar orbits that formed self-consistently (from the simulation by \citealt{map12}), rather than drawing stellar orbits with Monte Carlo sampling.}

The stars have semi-major axis $a$ ranging from $0.1 \pc$ to $0.4 \pc$ and mean eccentricity $\langle e \rangle = 0.3$ { (the eccentricity distribution is in excellent agreement with the data reported by \citealt{yel14})}. The disk has an initial opening angle of $\sim 7^\circ$. 

We simulate different inclinations $\theta_i$ between the infalling gas cloud and the stellar disk. To do this, we simply rotate the snapshots of the SPH simulation with respect to the plane of the stellar disk. We choose three different inclinations: $\theta_i = 10^\circ$ (run~A), $45^\circ$ (run~B), $90^\circ$ (run~C). The secular evolution of the orbital elements of a star in an axisymmetric potential strongly depends on the inclination between the orbital plane and the symmetry axis, we therefore expect different outcomes for  different inclinations. 
To compare our results, we also integrate the evolution of the stellar disk alone, without any infalling molecular gas cloud (run~D).
Table \ref{tab:ic} shows a summary of the runs presented in this paper.

\begin{figure}
  \centering
  \includegraphics[width=\linewidth]{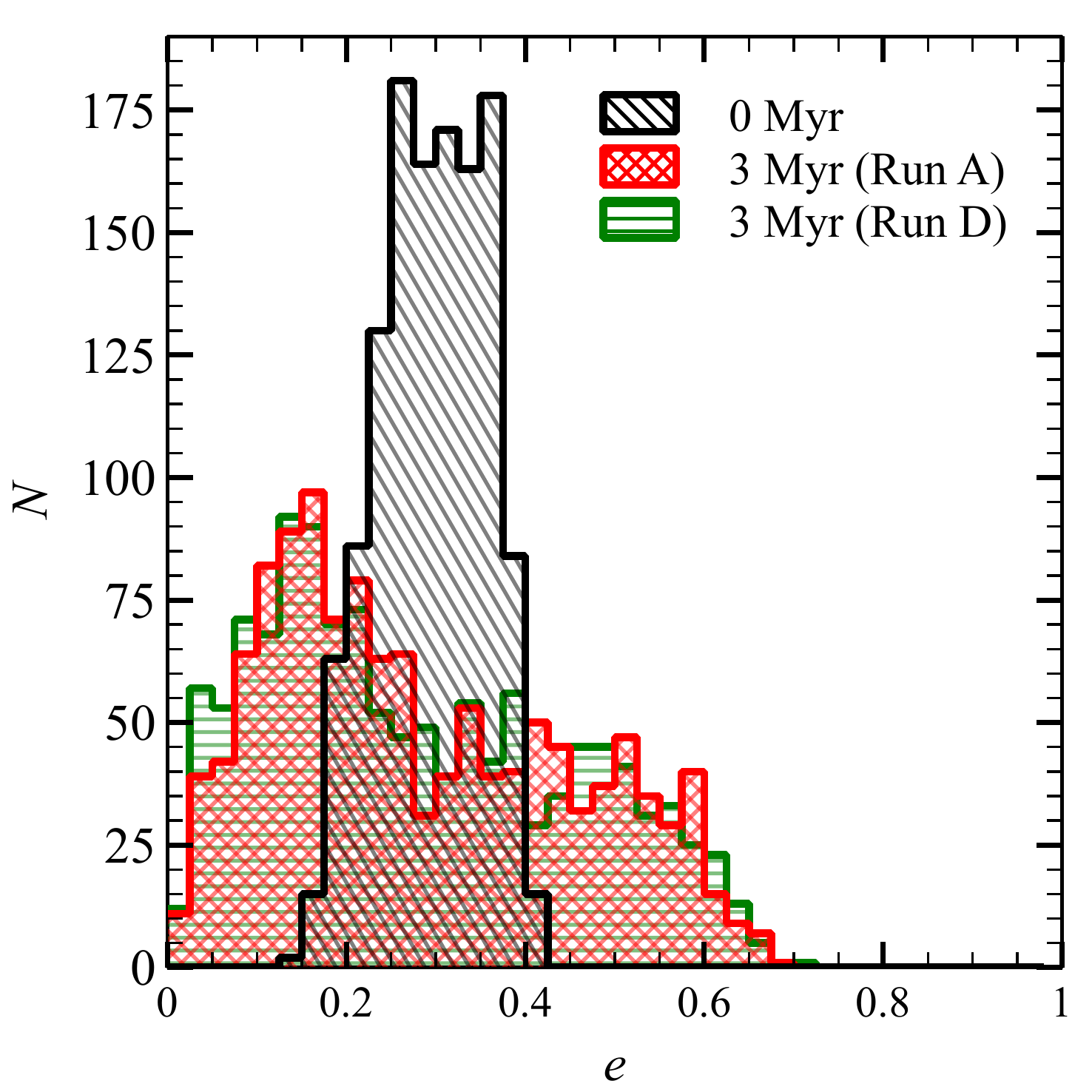}
  \includegraphics[width=\linewidth]{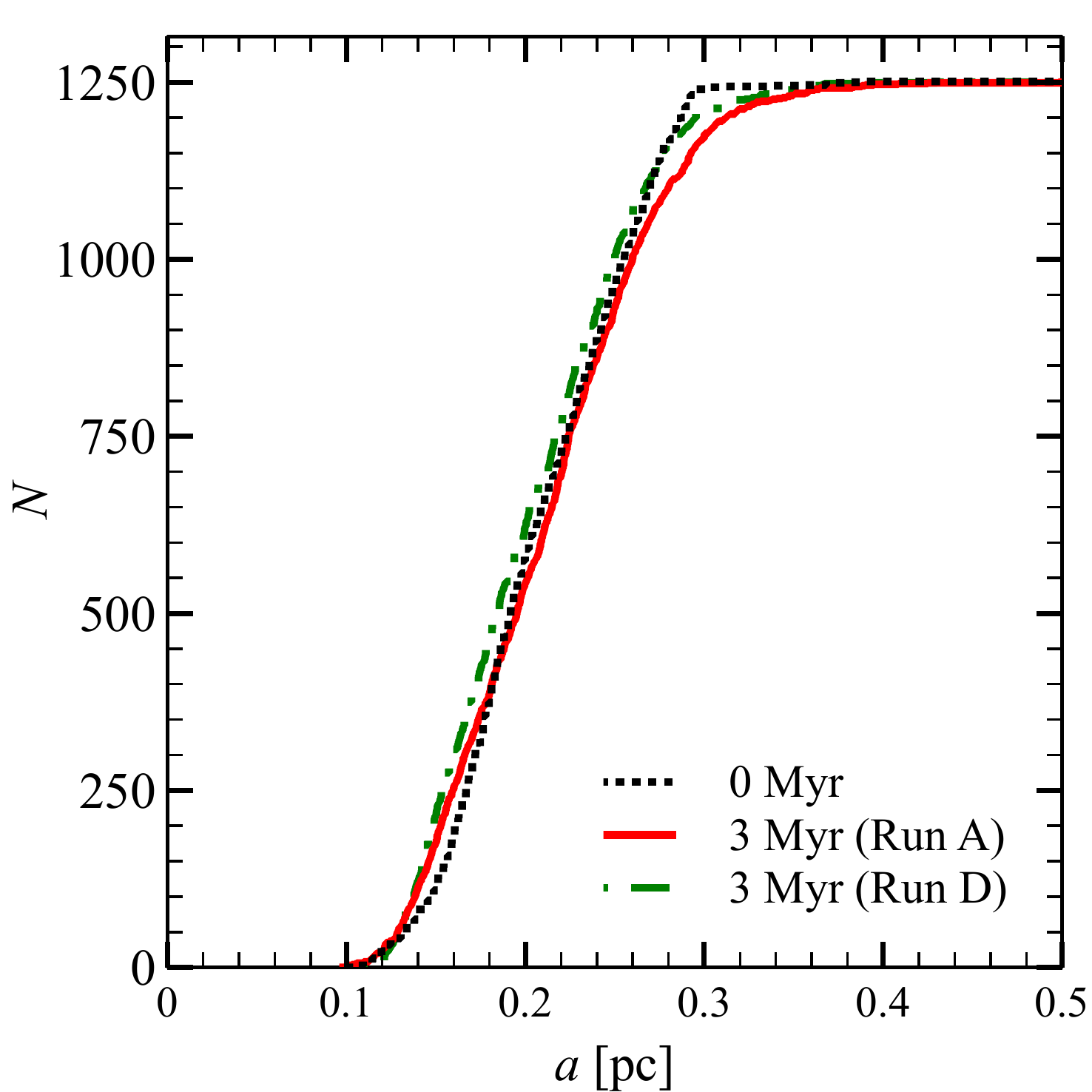}
  \caption{Top panel: eccentricity distribution of disk stars. Black hatched histogram: eccentricity distribution at the beginning of the integration. Red cross-hatched area: eccentricity distribution at $3 \myr$ in run~A ($\theta_i = 10^\circ$). Green horizontally-hatched histogram: eccentricity distribution at $3 \myr$ in run~D (no infalling gas cloud).
Bottom panel:  cumulative distribution of semi-major axis of disk stars. Black dotted line: distribution at the beginning of the integration. Red solid line: run~A ($\theta_i = 10^\circ$). Green dot-dashed line: run~D (no infalling gas cloud).
}
\label{fig:eadistro}
\end{figure}

{ We stop the simulations at 3 Myr, because this is the best observational estimate for the age of the young stars in the central parsec \citep{lu13}. Furthermore, after this time stellar mass changes and the energy input from supernovae and stellar winds become progressively more important.}

\section{Results}\label{sec:results}
\begin{figure*}
  \begin{center}
    \includegraphics[width=0.497\linewidth]{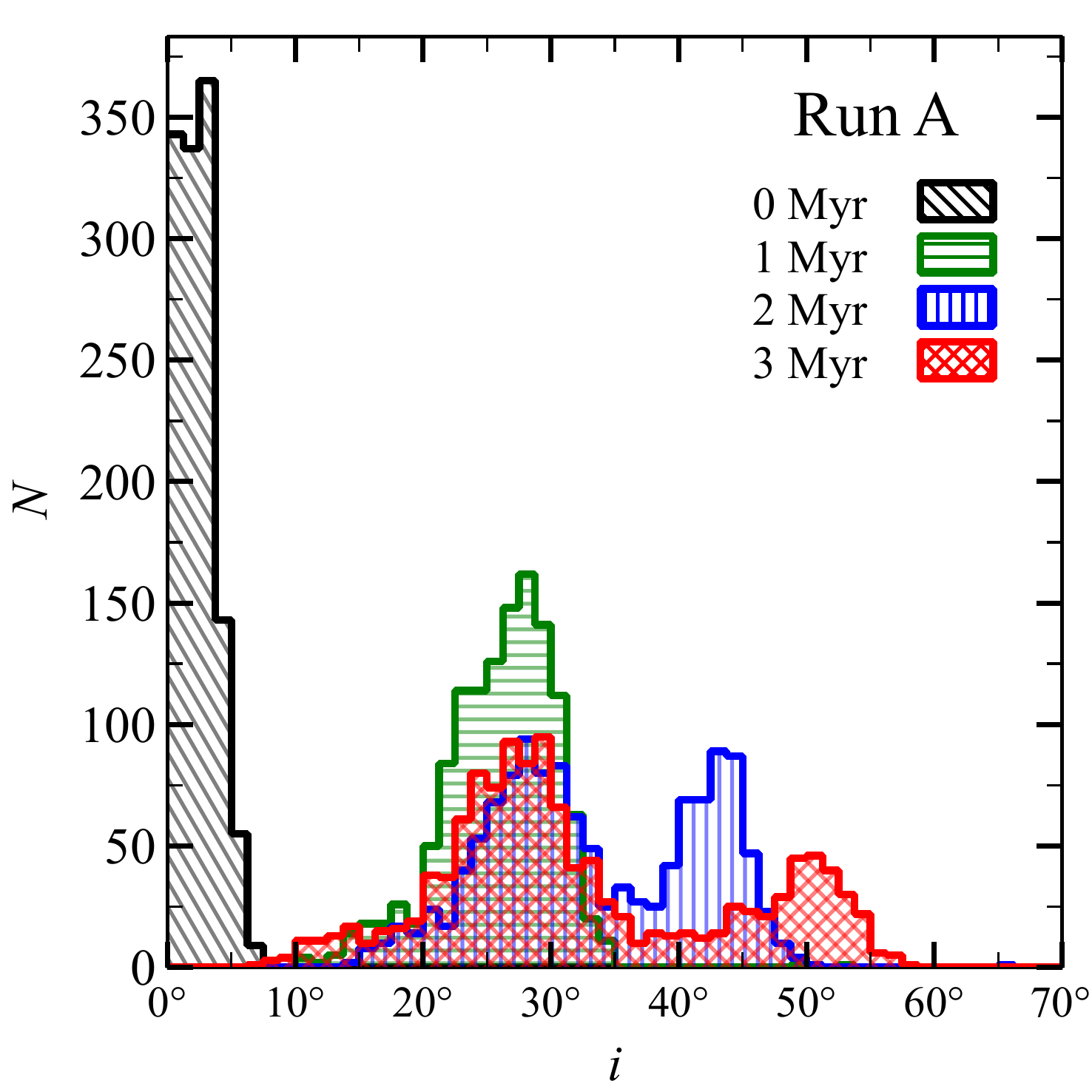}
    \includegraphics[width=0.497\linewidth]{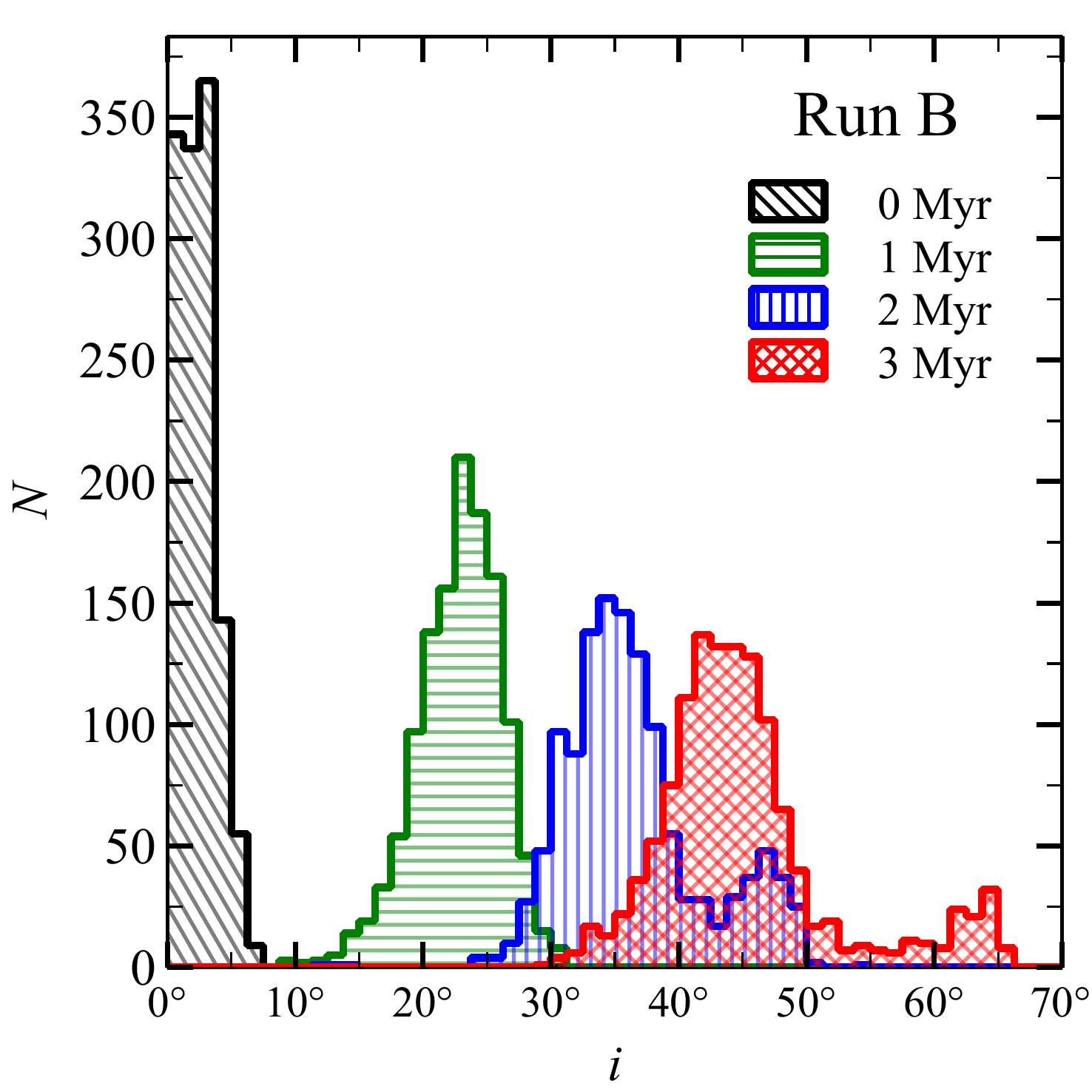}\\
    \includegraphics[width=0.497\linewidth]{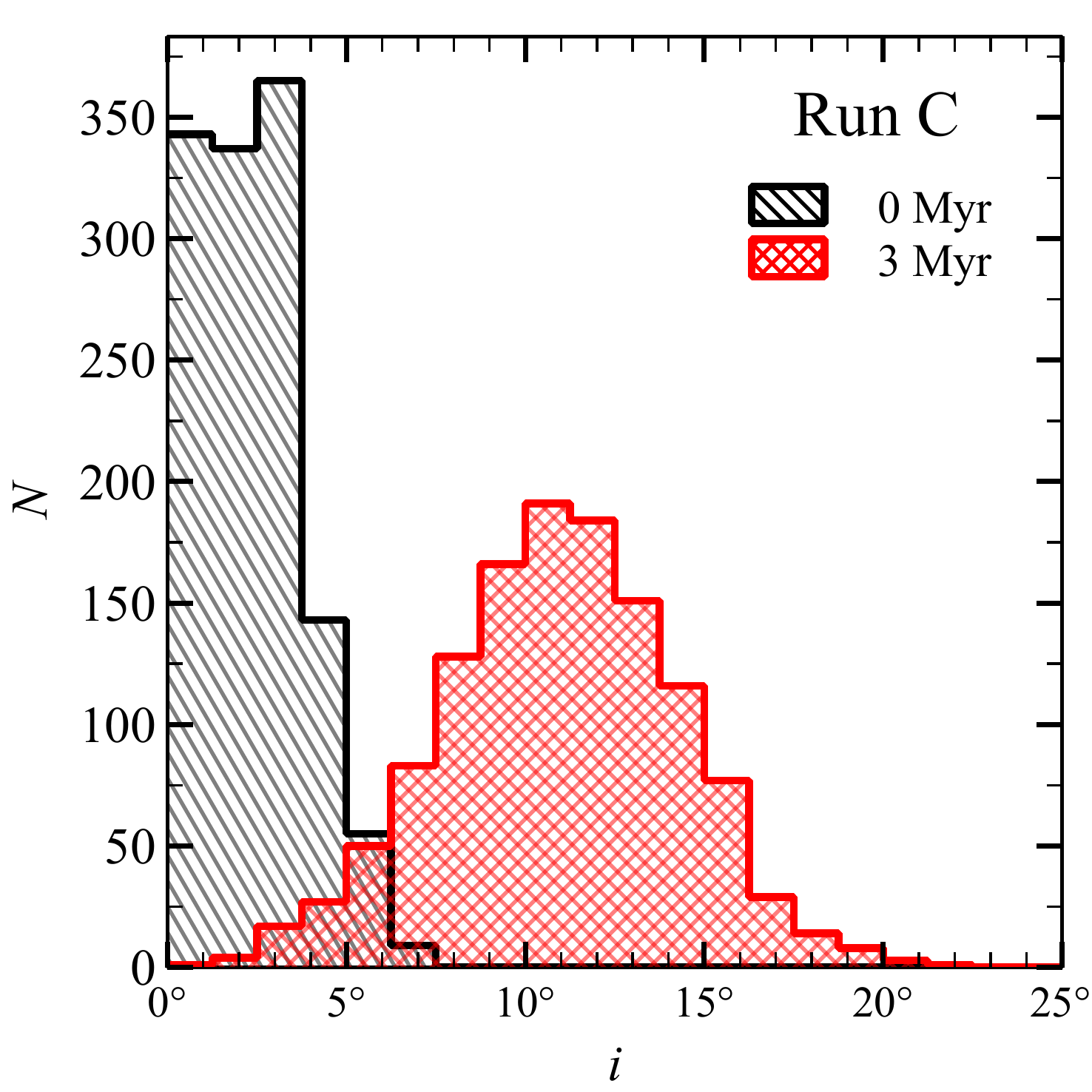}
    \includegraphics[width=0.497\linewidth]{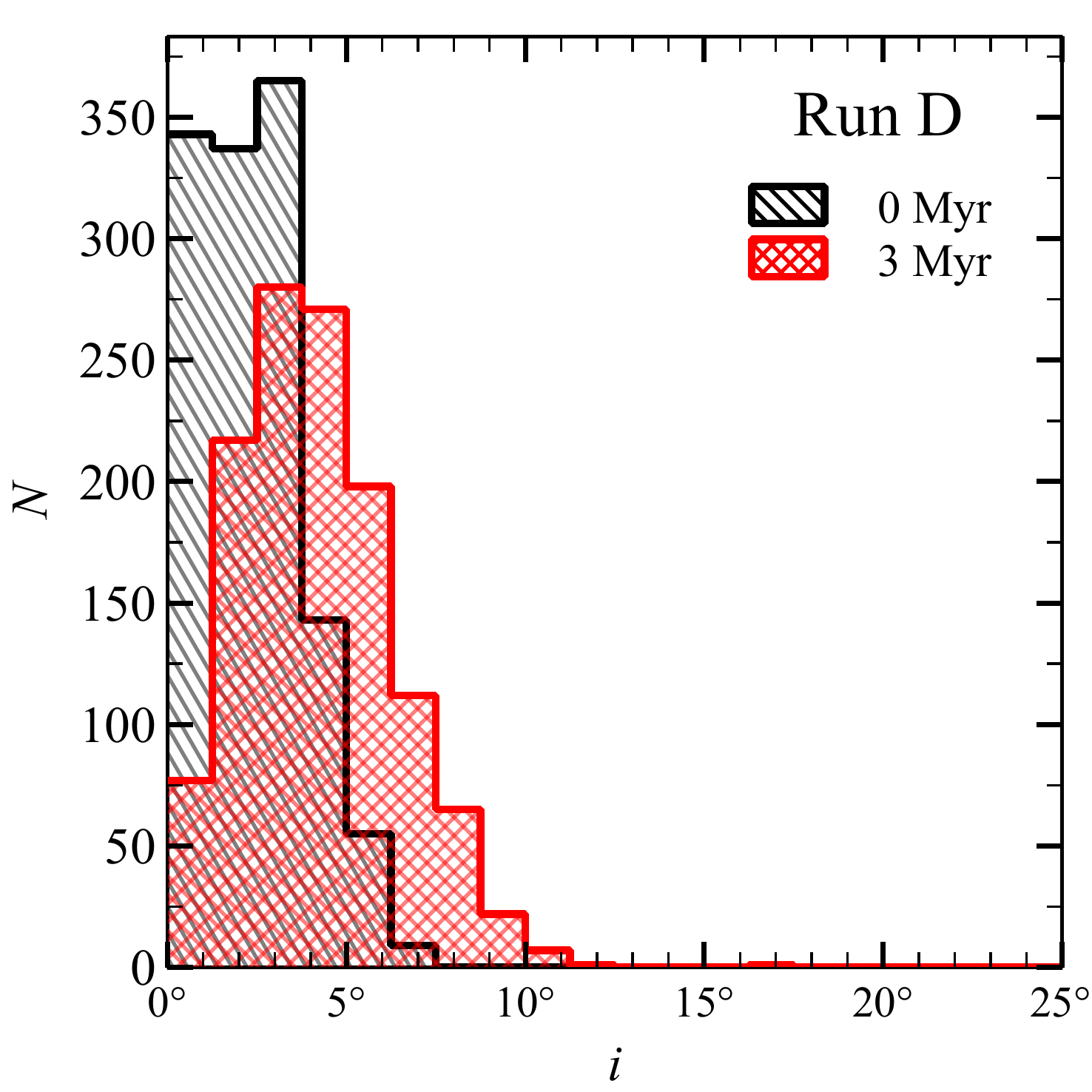}
  \end{center}
  \caption{Distribution of the inclination of stellar orbits. From top-left to bottom-right: runs~A, B, C and D. In all panels, black hatched histogram: distribution at the beginning of the integration; red cross-hatched histogram: distribution at $3 \myr$. In the top panels, green horizontally-hatched histogram: distribution at $1 \myr$; blue vertically-hatched area: distribution at $2 \myr$. In all panels, the inclination is measured with respect to the plane of the stellar disk at $0 \myr$.} 
    \label{fig:idistro}
\end{figure*}

In the first $1 \myr$, the gas cloud quickly inspirals towards the SMBH and settles down into a disk. Top (bottom) panel of Figure \ref{fig:gasmap} shows the projected density map of the gas ring seen face-on (edge-on) at $1 \myr$ { in run~A}. The formation of the gas ring is similar to the one described in section 3.1 of \citet{map13}. At $0.2 \myr$ the cloud is completely disrupted into several streamers. At $0.5 \myr$ the streamers begin to form a ring around the SMBH. The ring is progressively formed as the inspiraling streamers settle down around the SMBH. At $1.2 \myr$, each gas streamer has settled down and the gas ring is completely formed. The gas ring is actually composed of two concentric rings. The outer ring is composed of irregular and clumpy gas streamers and has an inner radius of $\sim 1.5 \pc$. The inner ring is warped and has a radius of $\sim 0.2-0.4 \pc$, similar to the stellar disk. The inner gas ring has a mass of $\sim 3\times 10^3\msun$, while the outer ring has a mass of $\sim 9\times 10^4\msun$. In addition, the inner ring is inclined by $\sim 24^\circ$ with respect to the outer ring, hence the stellar disk and the inner gas ring have a  mutual inclination $\theta_{\rm inner}$ (see Table \ref{tab:ic}).
{ This misalignment is due to the fact that the inner ring comes from low impact-parameter and low angular-momentum gas, which engulfs the SMBH during the first periapsis passage of the cloud, while the outer ring forms later, during
the subsequent periapsis passages of the disrupted cloud (and suffers from gravitational focusing and torques). 
 A detailed explanation is given in \citet{map15}.}

Once the gas ring has settled, it induces precession on the stellar disk, altering the stellar orbits. 
The evolution of eccentricity and semi-major axis distributions is similar in each run, both those including the infalling gas cloud (run A, B, C) and those which do not include gas (run~D). 
In particular, after $\sim 1.5 \myr$ the eccentricity distribution becomes bimodal, showing two peaks at $e \sim 0.15$ and $0.5$ (Figure \ref{fig:eadistro}, top panel).

The semi-major axis distribution does not change significantly throughout the simulations (Figure \ref{fig:eadistro}, bottom panel), indicating that in our model two-body relaxation of orbital energy is inefficient. This result is in agreement with \citet{sub14}, { who find} that two-body relaxation is inefficient in changing the semi-major axis distribution if $\langle e^2 \rangle^{1/2} \gtrsim 0.3$.

On the other hand, the torques exerted by the gas ring strongly affect the orbital inclinations. In Figure \ref{fig:idistro} we show the inclination distribution at $3 \myr$ in each run, compared with the initial conditions. 
In the case of Run A (Figure \ref{fig:idistro}, top-left panel), the inclinations of the stars change significantly from the initial one, showing two peaks at $\sim 27^\circ$ and $50^\circ$ degrees from the initial disk inclination. This indicates that the whole disk changed its orientation during the simulation.

In run B (Figure \ref{fig:idistro}, top-right panel), there is less spread in the inclinations than in run A and the inclination distribution at $3 \myr$ shows a main peak at $\sim 42^\circ$. As in run A, this means that the disk has completely changed its orientation with respect to the initial configuration. In run C the inclination distribution at $3 \myr$ is peaked at $\sim 12^\circ$ and shows little spread.

In the case of Run D (Figure \ref{fig:idistro}, bottom-right panel), the disk is unperturbed and the inclinations do not change as significantly as in the other runs.  At $3 \myr$ the inclination distribution is broader than the initial one, but the disk has preserved its initial orientation.

\begin{figure}
  \centering
  \includegraphics[width=\linewidth]{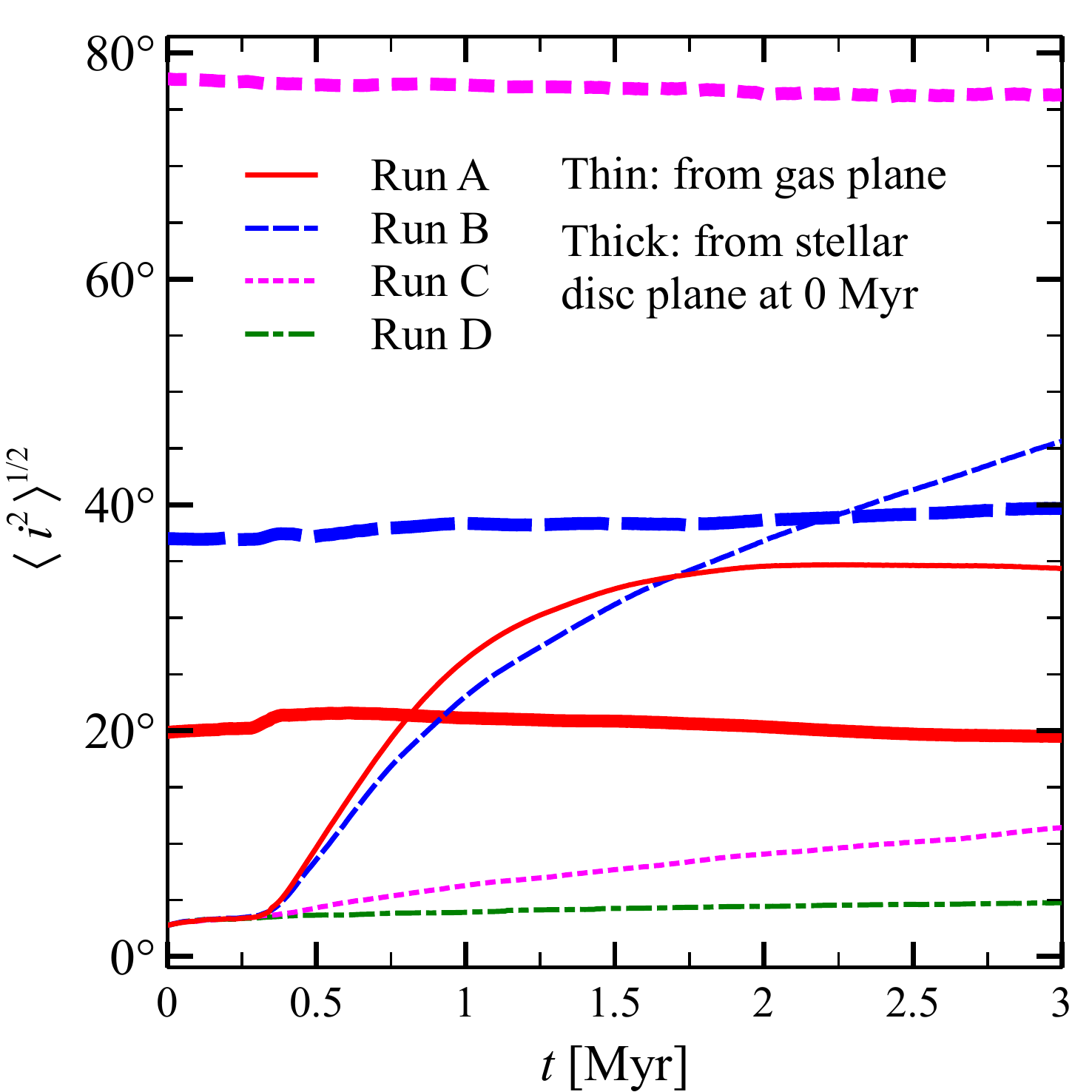}
  \caption{Root mean square inclination of the stellar disk as a function of time. Thin lines: inclination measured with respect to the plane of the stellar disk at $0 \myr$. Thick lines: inclination measured with respect to the plane of the gas inner ring at $0.5 \myr$. Red, solid lines: run A ($\theta_i = 10^\circ$). Blue, dashed lines: run B ($\theta_i = 45^\circ$). Magenta, dotted lines: run C ($\theta_i = 90^\circ$). Green, dot-dashed line: run D (no infalling gas cloud).}
\label{fig:rmsi}
\end{figure}

In Figure \ref{fig:rmsi} we show the evolution of the root mean square inclination $\langle i^2 \rangle^{1/2}$ of the stars.
In runs A, B and C, $\langle i^2 \rangle^{1/2}$ begins to increase after $0.75 \myr$. 
In run A $\langle i^2 \rangle^{1/2}$ does not increase above $35^\circ$, while in run B it continues to increase. In run C the inclination increases more slowly, while in run D the evolution of $\langle i^2 \rangle^{1/2}$ is negligible compared to the other runs.

The large spread in the inclination distribution of runs A and B indicates that the disk loses its coherence. In order to quantify the number of stars that are displaced from the disk, we define the disk membership criterion in a similar way as the one used by \citet{haa11a}. In particular, we assume that a star belongs to the disk if its angular momentum deviates from the mean normalized angular momentum by less than $20^\circ$. We recompute the mean normalized angular momentum whenever a star is rejected as a disk member, until the number of disk members does not change anymore. 

\begin{figure}
  \centering
  \includegraphics[width=\linewidth]{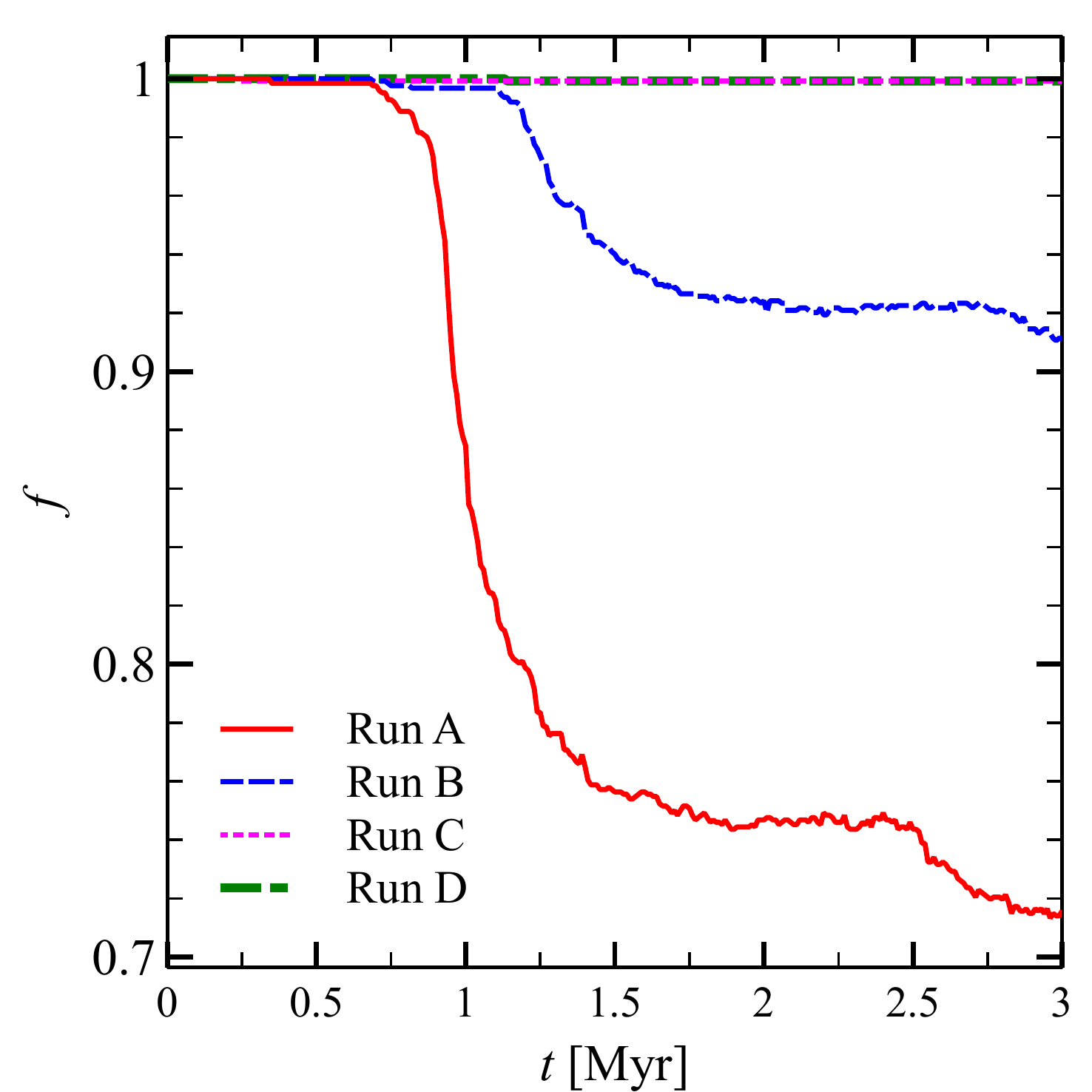}
  \caption{Fraction of stars with angular momentum deviating from the total angular momentum of the stellar disk by less than $20^\circ$ as a function of time. Red, solid line: run A ($\theta_i = 10^\circ$). Blue, dashed line: run B ($\theta_i = 45^\circ$). Magenta, dotted line: run C ($\theta_i = 90^\circ$). Green, dot-dashed line: run D (no infalling gas cloud). The magenta and green lines overlap.}
\label{fig:int}
\end{figure}

The number of disk members for different runs as a function of time is shown in Figure \ref{fig:int}. While the number of disk members remains constant in runs C and D, in the other runs it begins to decrease after $0.75 \myr$. In particular, the number of disk members decreases abruptly at $\sim{}$ 1 Myr and $\sim{}$ 1.25 Myr in runs~A and B, respectively. This abrupt change is a consequence of how disk membership is defined, indicating that most of the displaced stars cross the $20^\circ$ threshold approximately at the same time.
We find that at $3 \myr$ the number of disk members decreased by $\approx{}30\%$, $10\%$ in run A and B, respectively. 

\begin{figure}
  \centering
  \includegraphics[width=\linewidth]{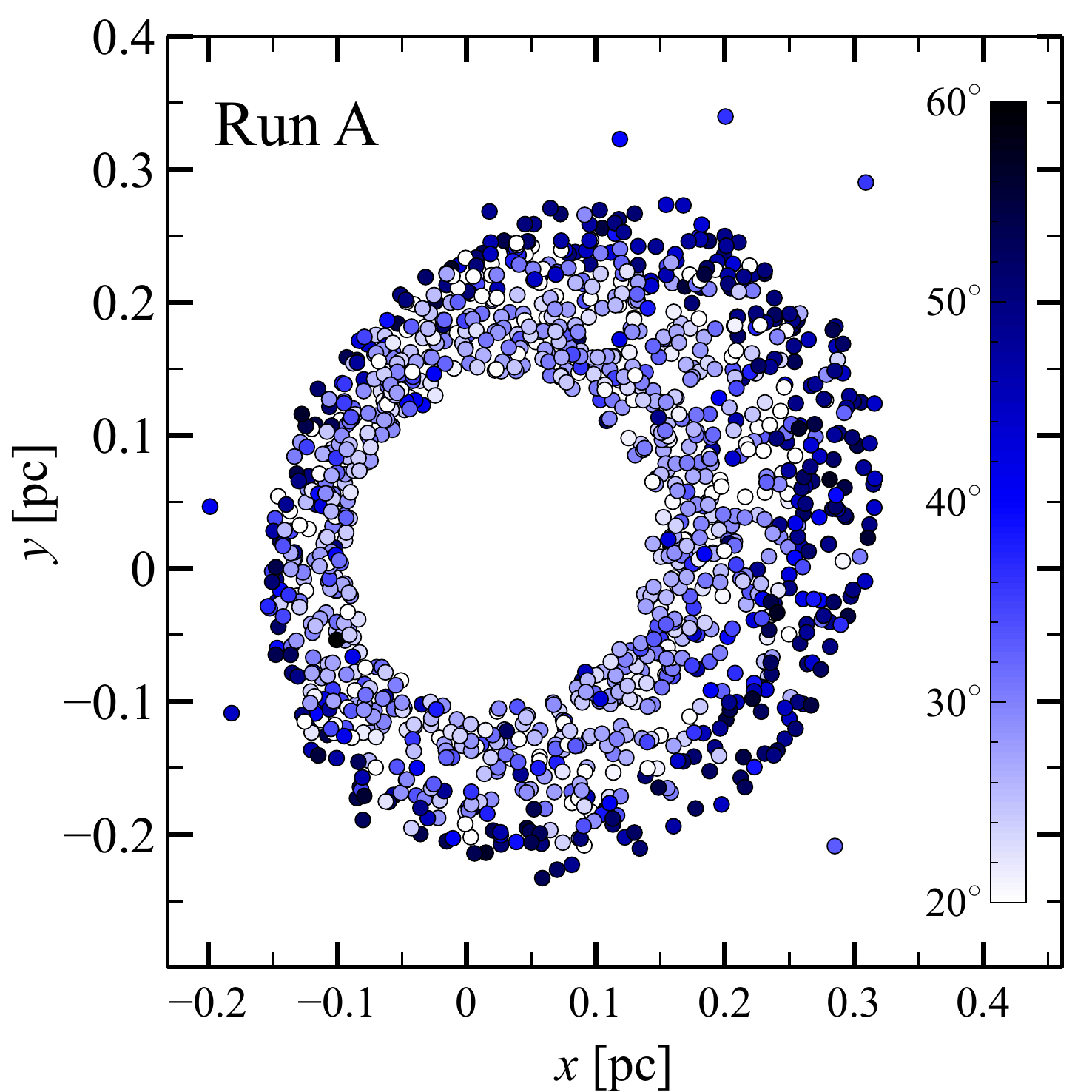}
  \caption{Position of disk stars at $0 \myr$, projected along the normal to the disk. The circles are colored according to the star inclinations at $3 \myr$ for run A ($\theta_i = 10^\circ$). The inclinations are measured with respect to the normal to the stellar disk at $0 \myr$ and range from $20^\circ$ to $60^\circ$.}
\label{fig:xyA}
\end{figure}

Figure \ref{fig:xyA} shows the initial position of the stars, with the colors indicating the inclination of the star orbit at $3 \myr$ in run~A. The color gradient indicates that the stars initially on outer orbits reach an higher inclination than the stars on inner orbits.

\begin{figure}
  \centering
  \includegraphics[width=\linewidth]{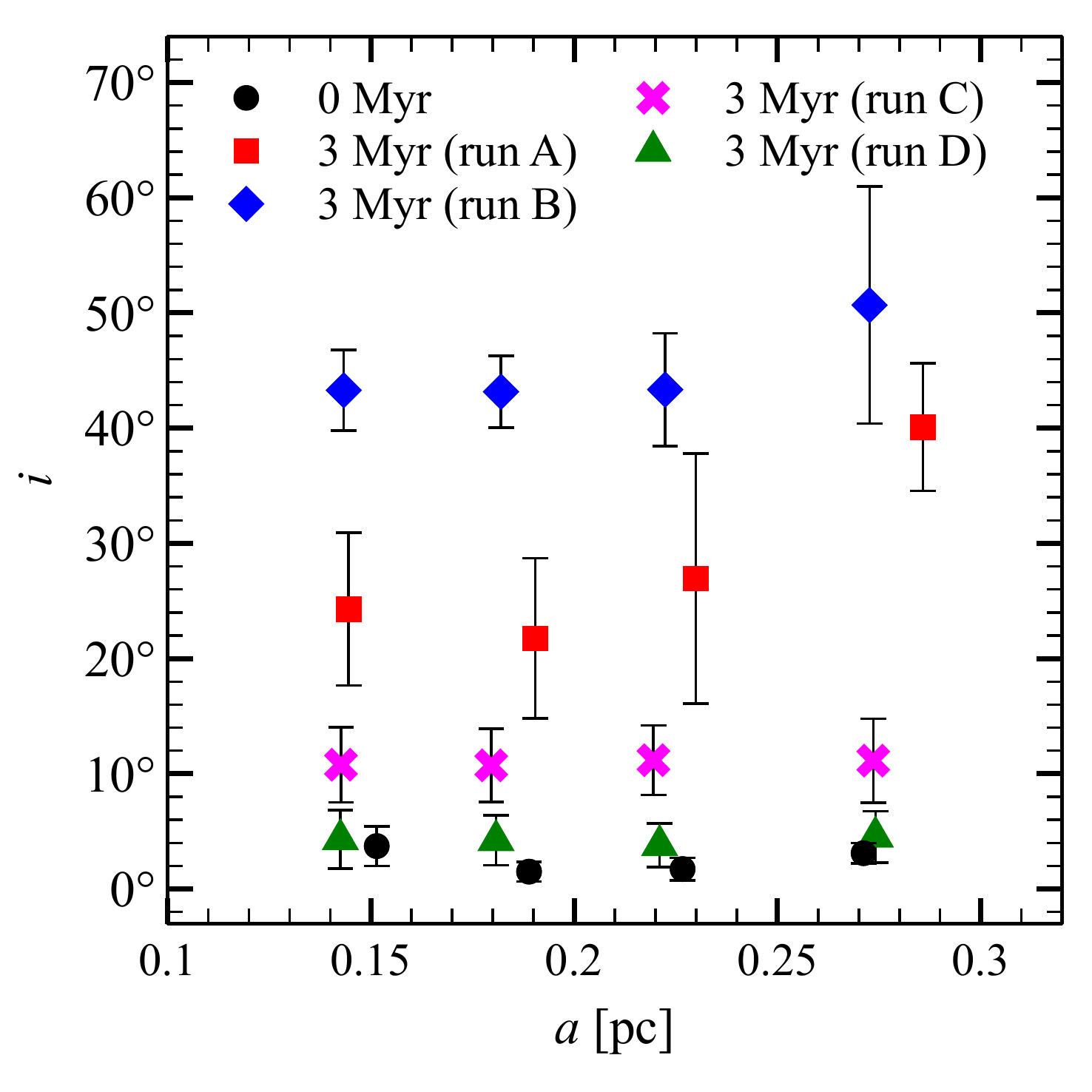}
  \caption{Inclination as a function of the semi-major axis binned in four radial bins, so that each bin contains an equal number of stars. Black circles: Inclination at $t=0 \myr$. Red squares: inclination at $3 \myr$ for run A ($\theta_i = 10^\circ$). Blue diamonds: inclination at $3 \myr$ for run B ($\theta_i = 45^\circ$). Magenta crosses: inclination at $3 \myr$ for run C ($\theta_i = 90^\circ$). Green triangles: inclination at $3 \myr$ for run D (no infalling gas cloud). The inclination is measured with respect to the normal to the stellar disk at $0 \myr$. Each point is the average inclination per bin, while the error bars show the standard deviation for each bin.} 
\label{fig:iascatter}
\end{figure}

This trend is confirmed by Figure \ref{fig:iascatter}, that shows the average inclination as function of semi-major axis for run A, B and C at $3 \myr$. In the initial disk the inclinations are of $\sim 7^\circ$, regardless of the semi-major axis. At $3 \myr$, the average inclination has increased by $30^\circ{}-50^\circ$, depending on the run, and the individual inclinations have spread, as shown by the error bars. Moreover, in runs A and B the average inclination is higher for increasing semi-major axis. The origin of these differences between runs is discusses in the next section.

\section{Discussion}\label{sec:discussion}
\subsection{The importance of the inner gas ring}
In our numerical model, the dynamics of the stellar disk is driven by two processes: two-body relaxation in the stellar disk and precession induced by the external potentials. The external potentials are the analytical cusp, which induces precession of the periapsis and the gas ring. 
The gas ring is irregular and clumpy, however we can consider its potential to be axisymmetric as a first approximation.
Such axisymmetric potential induces precession of the mean orbital elements of a single star on a timescale given by the Kozai-Lidov timescale: 
\begin{equation}\label{eq:tk}
T_{\rm K}=\frac{M_{\rm SMBH}}{M_{\rm CNR}} \frac{R_{\rm CNR}^3}{\sqrt{GM_{\rm SMBH} a^3}}
\end{equation}
where $M_{\rm CNR}$ and $R_{\rm CNR}$ are the mass and radius of the gas ring, $G$ is the gravitational constant and $a$ is the semi-major axis of the star. The equations of motion of the mean orbital elements also depend on eccentricity $e$ and argument of periapsis $\omega$ \citep[e.g. see][]{sub09}.

Since the gas ring is actually composed of two concentric rings, each component induces precession on the stellar disk on a different Kozai-Lidov timescale.
The inner ring is less massive than the outer ring by a factor of $\sim 30$, but has a $\sim 5$ times smaller radius. Due to the Kozai-Lidov timescale dependence $T_{\rm K} \propto M_{\rm CNR}^{-1}\,R_{\rm CNR}^3$, the inner ring induces precession on a shorter timescale than the outer ring. We find $T_{\rm K} \sim 2 \myr$ for the inner ring and $T_{\rm K} \sim 8 \myr$ for the outer ring. Therefore, we expect that the precession of stars will be mainly driven by the inner ring, rather than by the outer one. 

\begin{figure}
  \centering
  \includegraphics[width=\linewidth]{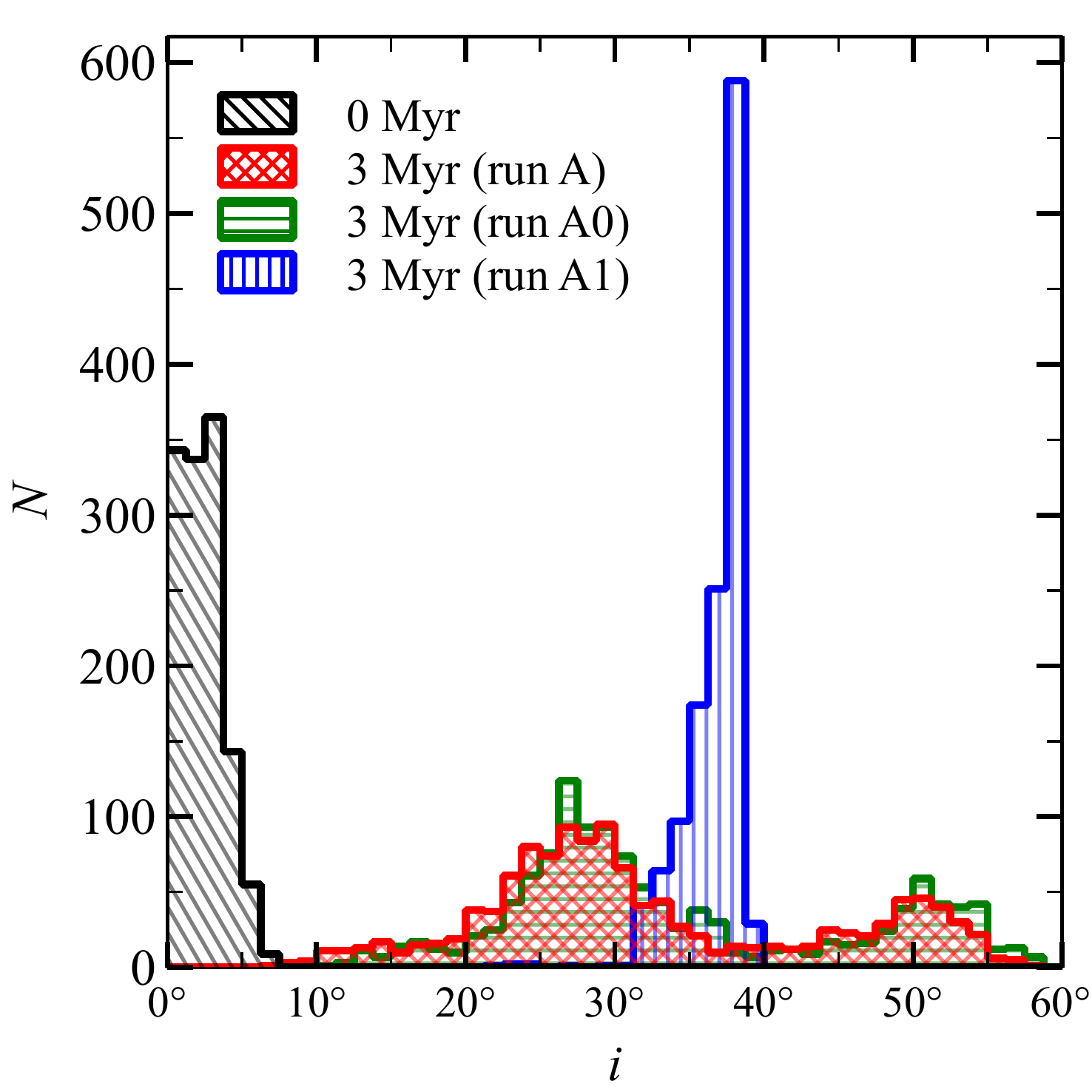}
  \caption{
Inclination distribution of disk stars. Black, hatched area: $0 \myr$. Red, cross-hatched area: run A at $3 \myr$ ($\theta_i = 10^\circ$). Green, horizontal-hatched area: run A0 at $3 \myr$ ($\theta_i = 10^\circ$, no outer gas ring). Blue, vertical-hatched area: run A1 at $3 \myr$ ($\theta_i = 10^\circ$, massless stars). The inclination is measured with respect to the plane of the stellar disk at $0 \myr$.}
\label{fig:idistrobackup}
\end{figure}

To check the importance of the outer ring, we run a simulation with the same initial conditions as run A, but removing the gas particles with radius $a > 0.9 \pc$ from the snapshots of the SPH simulation with $t > 0.5 \myr$. In this way, we remove the outer ring from the simulation without affecting the evolution of the inner ring. In Figure \ref{fig:idistrobackup}, we compare the inclination distribution of this run, named A0, with that of run A. The differences between the two runs are negligible, thus the stellar disk is not affected by the outer gas ring, and its evolution is driven mainly by the inner gas ring.

\begin{figure*}
  \begin{center}
    \includegraphics[width=0.497\linewidth]{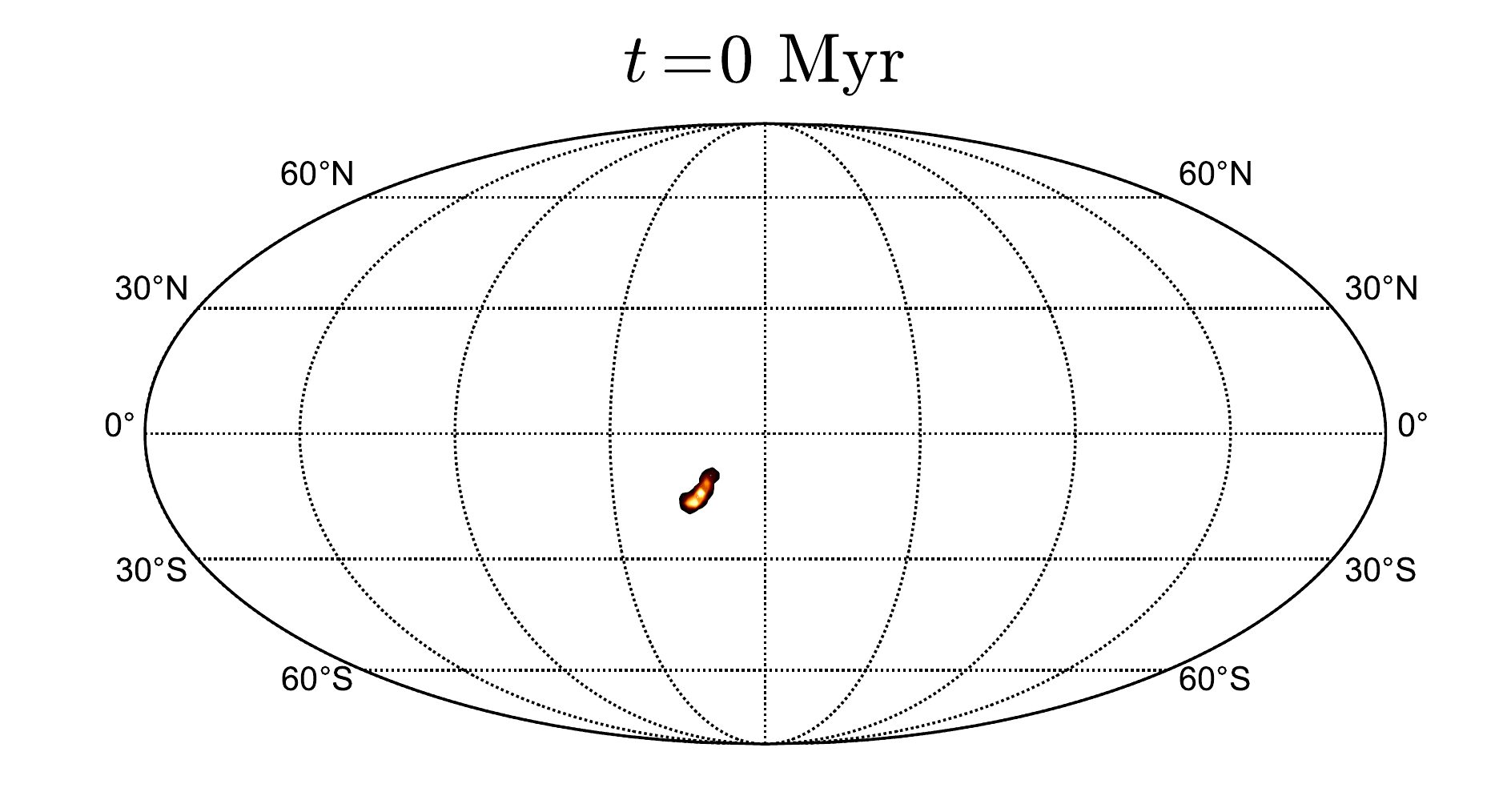}
    \includegraphics[width=0.497\linewidth]{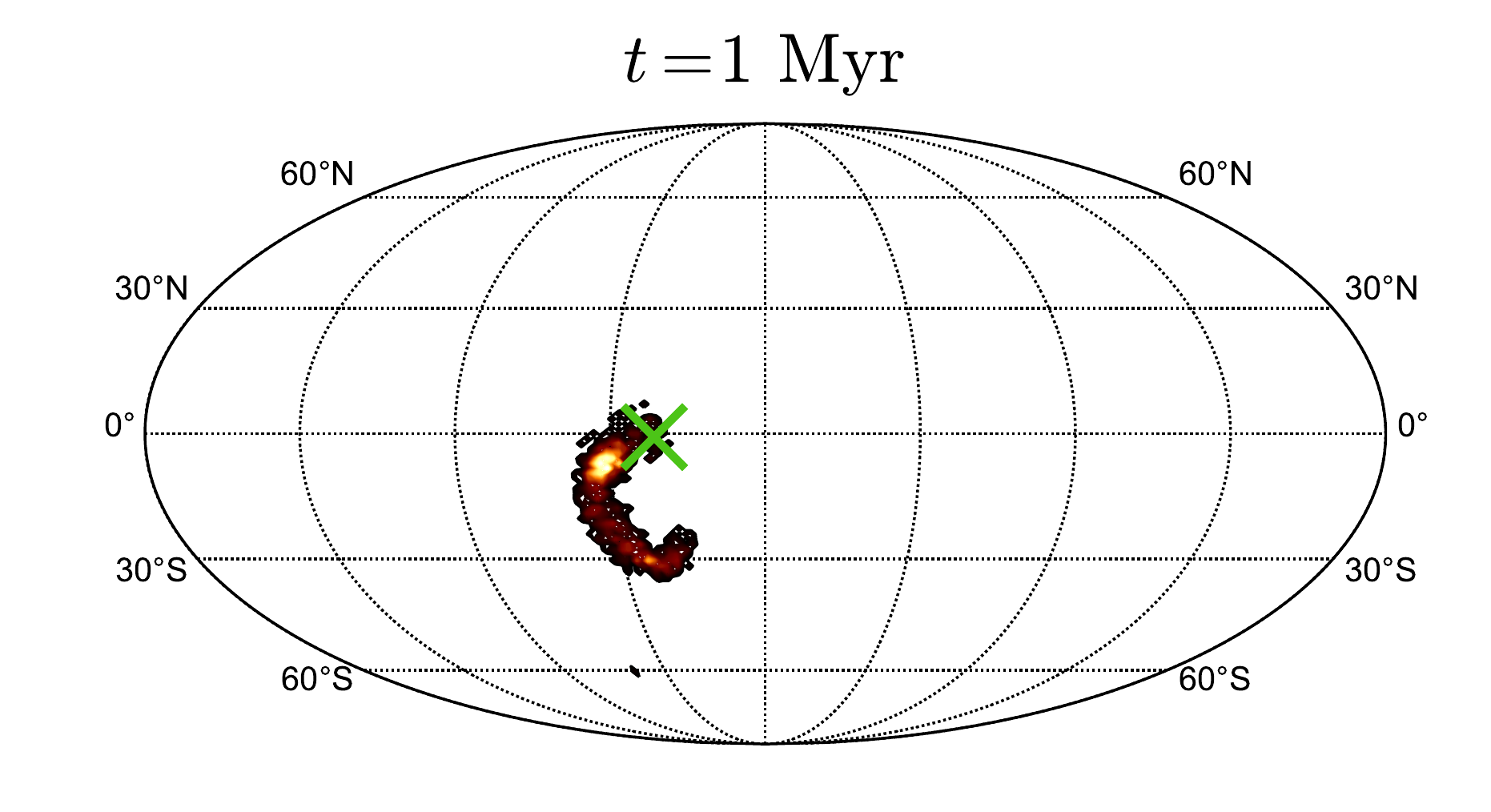}\\
    \includegraphics[width=0.497\linewidth]{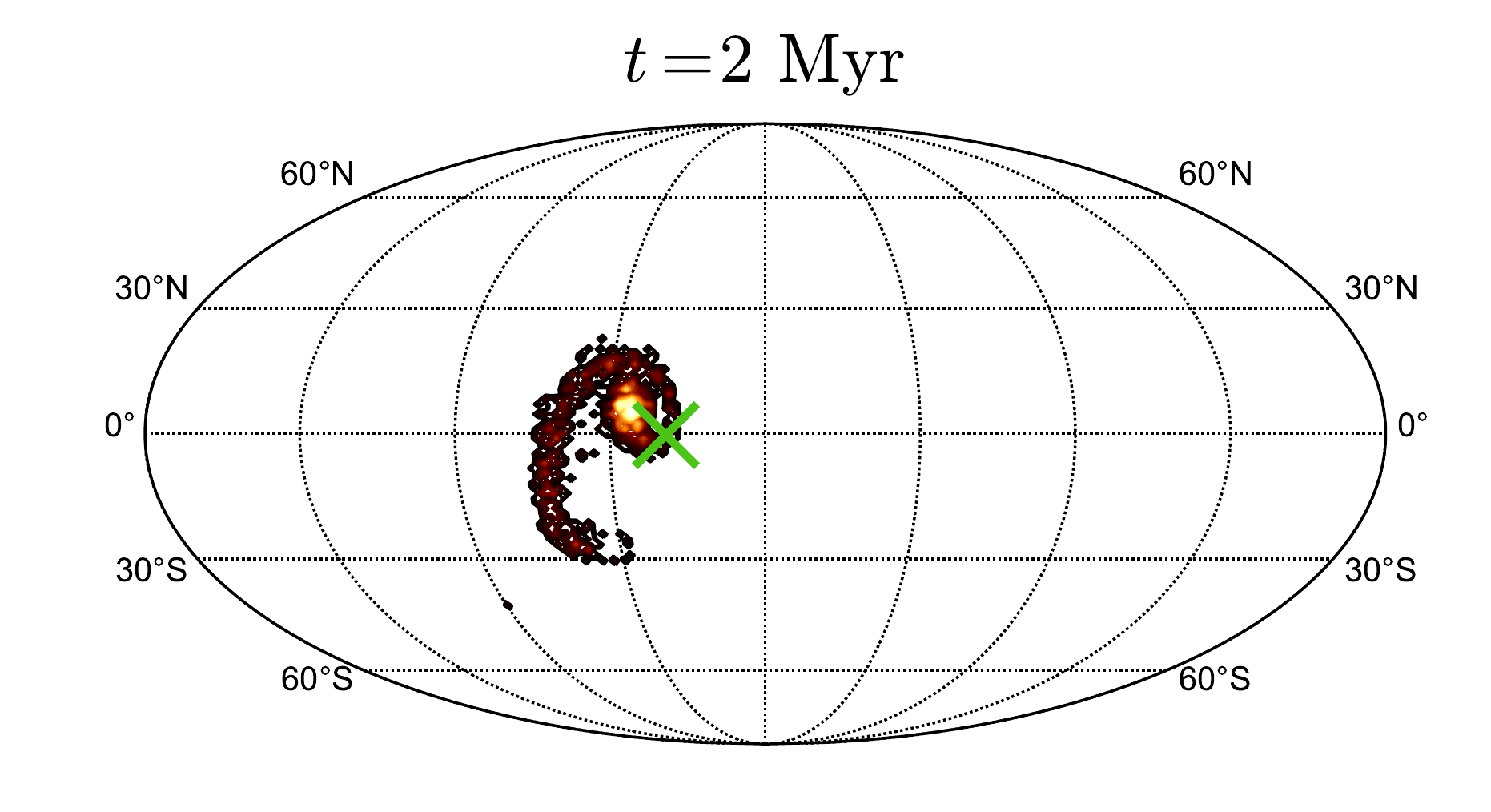}
    \includegraphics[width=0.497\linewidth]{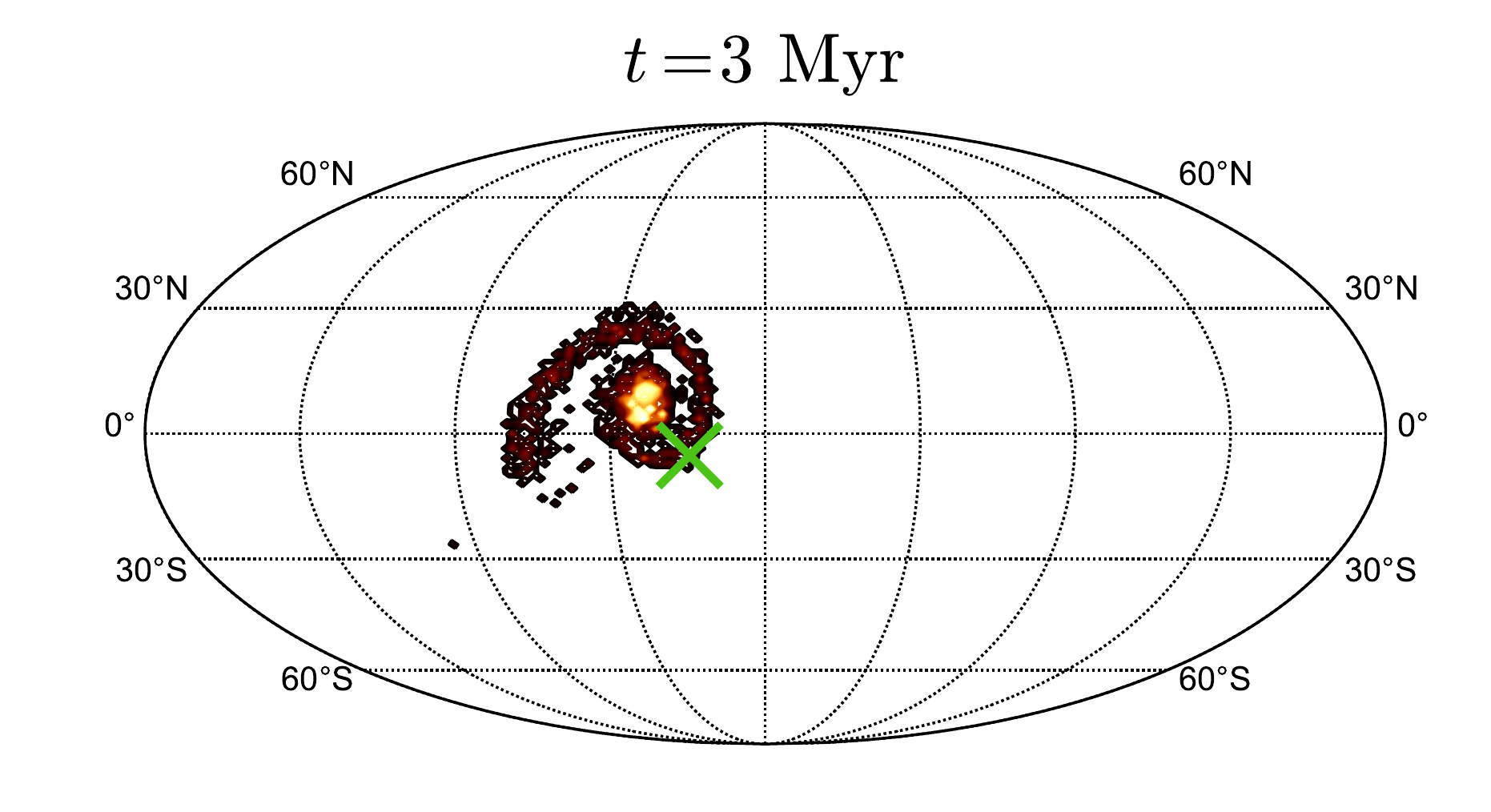}
  \end{center}
  \caption{Density of normal vectors to the stellar orbits at different times for run~A. From top-left to bottom-right: $t = 0, 1, 2$ and $3 \myr$. The green cross indicates the total angular momentum vector of the inner gas ring (not yet formed at $t = 0 \myr$). Projected using the  Mollweide projection. 
}
    \label{fig:maps}
\end{figure*}

Figure \ref{fig:maps} shows the time evolution of the density of normal vectors to the stellar orbits in run~A. The normal vectors rotate about the angular momentum vector of the inner gas ring (marked with a green cross), losing coherence and forming a spiral-shaped tail. The normal vectors that form the spiral-shaped tail in \ref{fig:maps} correspond to the secondary peak at $50^\circ$ in the top-left panel of Figure \ref{fig:idistro}. { A similar spiral-shaped pattern was found by \citet{loc09}, who simulated the interaction between two mutually inclined stellar disks \citep[see fig.~2 of][]{loc09}.} The formation of the tail is due to the combined effect of two-body relaxation and $\Omega$ precession and will be discussed in Section~\ref{sec:twob}. 

In Section~\ref{sec:results} (Figures~\ref{fig:idistro}, \ref{fig:rmsi} and \ref{fig:int}), we showed  that the inclination distribution changes faster in runs~A and B than in run~C. The reason is that the precession induced by an axisymmetric potential strongly depends on the angle  between the individual stellar orbits and the gas ring. The timescale for precession of the longitude of the ascending node $\Omega$ scales as $\cos^{-1}{\theta}$. In fact, in run C, $\theta_{\rm inner}\simeq 77^\circ$ and the $\Omega$ precession is strongly suppressed. As a consequence, the rate of increase of stellar inclinations in run C is much smaller than in run A and B, while in run A ($\theta_{\rm inner} \simeq 22^\circ$) is larger than in run B ($\theta_{\rm inner} \simeq 37^\circ$) (Figure \ref{fig:rmsi}).

However, the increase of stellar inclination in run A halts after $1.5 \myr$. This is due the fact that the changes in inclination are related to the changes in the longitude of the ascending node $\Delta \Omega$, which are limited in the range $0-2\pi$. The maximum inclination that can be achieved is $i= 2\,\theta_{\rm inner}$, when $\Delta \Omega = \pi$. Since in run A $\theta_{\rm inner} \simeq 20^\circ$, the inclination does not increase past $i\simeq 40^\circ$. After $\Delta \Omega > \pi$ the inclination should decrease, while in run A it remains approximately constant. This can be due to the fact that these predictions hold in the approximation of a static, axisymmetric potential, while the inner gas ring in our simulations consists of particles that are gravitationally interacting. 
Two-body relaxation might also play a role in this.

The inner gas ring in our simulations has a density of $10^6\,\rm cm^{-3}$ and a temperature of $100-500\,\rm K$, indicating that is composed of warm, neutral gas.  
\citet{jac93} estimated the presence of $\gtrsim 300\msun$ of dense neutral gas associated to the ionized gas within the cavity of the CNR.
This estimate is one order of magnitude smaller than the mass of the inner ring in our simulations, but it is uncertain and only poses a lower limit to the amount of gas present, since it does not account for ionized and molecular gas.
More recent far-infrared observations suggest that shocks and/or photodissociation dominate the heating of hot molecular gas in central cavity \citep{goi13}. Therefore, a more accurate modeling of the central cavity would require a better treatment of ionization and radiative transfer.
On the other hand, our results suggest that the kinematics of the CW disk (and of the other young stars) can give us constraints on the gas mass in the central cavity.

\begin{figure}
  \centering
  \includegraphics[width=\linewidth]{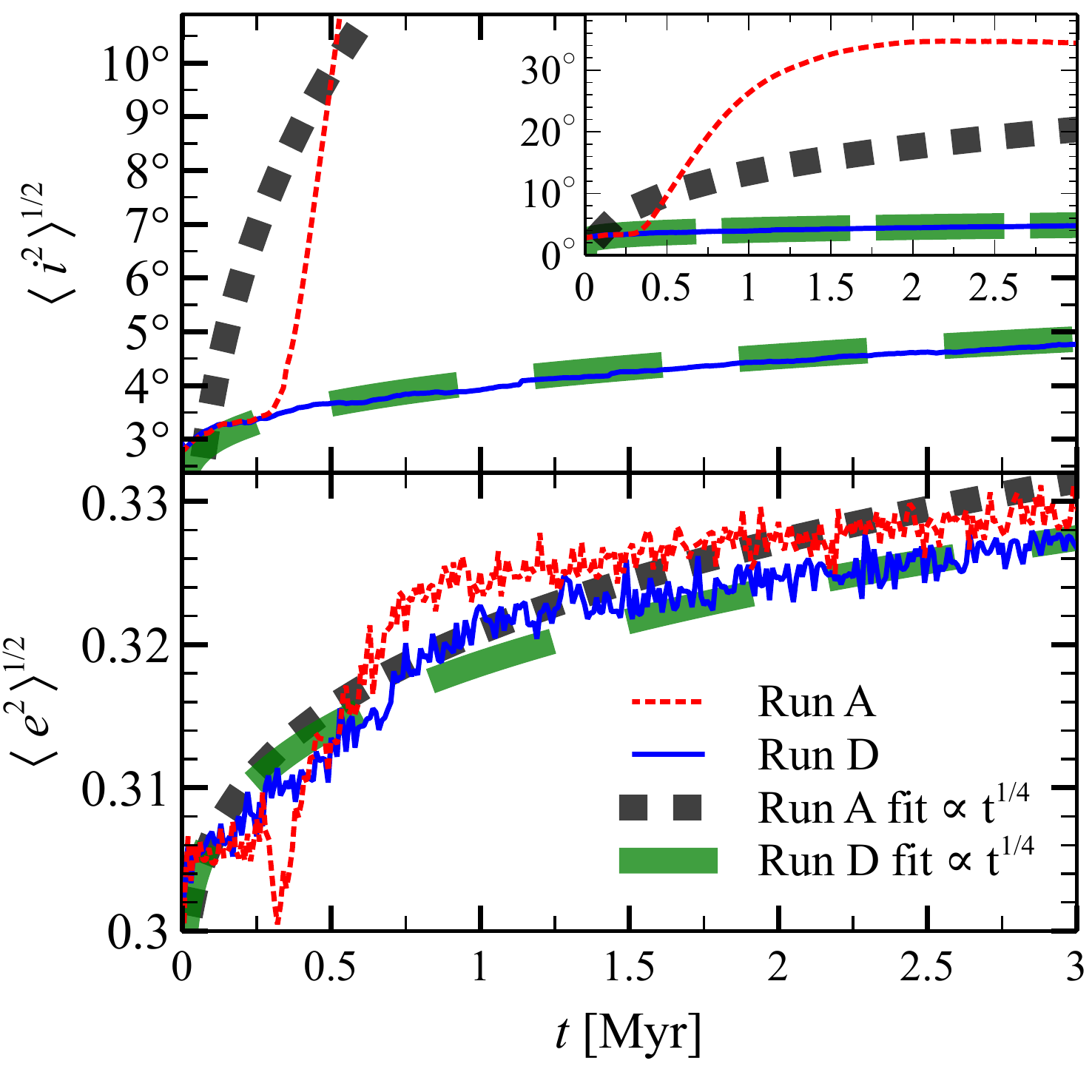}
  \caption{Root mean squared inclination (top) and eccentricity (bottom) as a function of time. Thin blue solid line: rms values for run D (no infalling gas cloud). Thin red dotted line: rms values for run A ($\theta_i = 10^\circ$). The thick lines are fit functions $\propto t^{1/4}$, dashed green for run D and dotted black for run A. In the inset: same as the top panel, but zoomed-out to $40^\circ$. 
}
\label{fig:rmsie}
\end{figure}

\subsection{Two-body relaxation}\label{sec:twob}
It has been demonstrated that the apsidal precession induced by the spherical cusp can suppress Kozai-Lidov oscillations \citep{cha09,loc09}. In fact, we do not find evidences of Kozai-Lidov cycles either in inclination or eccentricity. Nonetheless, we find changes in the inclination and eccentricity distributions in our simulations (Fig.~\ref{fig:eadistro}, Fig.~\ref{fig:idistro}). The combined effect of two-body relaxation and eccentric disk instability accounts for changes in eccentricity.

The eccentricity becomes bimodal after $1.5 \myr$ (Figure \ref{fig:eadistro}, top panel). This is due to the eccentric disk instability described by \citet{mad09}, according to which the precession induced by a stellar cusp drives coherent torques that amplify deviations of individual orbital eccentricities from the average. In fact, we find the same bimodal distribution whether or not we include the gas ring. This result is in agreement with \citet{gua12}, who find the same bimodal distribution evolving a similar stellar disk in the potential of a stellar cusp. 

While eccentric disk instability accounts for the bimodality of the eccentricity distribution, two-body relaxation accounts for its overall evolution.
According to two-body relaxation theory of thin disks, the root mean square eccentricity and inclination should grow over time as \citep{ste00,sub14}: 
\begin{equation}
\langle i^2 \rangle^{1/2} \propto \langle e^2 \rangle^{1/2} \propto t^{1/4}
\end{equation}

Figure \ref{fig:rmsie} shows the evolution of $\langle e^2 \rangle^{1/2}$ and $\langle i^2 \rangle^{1/2}$ along with the analytical predictions. It is apparent that the eccentricity evolution is well fit by the analytical predictions for both run A and D. This seems to indicate that the evolution of the eccentricity is predominantly driven by two-body relaxation. 

This is not the case for the evolution of inclinations. While the root mean square inclination of run D is consistent with the analytical predictions of two-body relaxation, the fit is inconsistent for run A, B and C. This is due to torques exerted by the gas ring onto the stellar disk. In fact, while the spherical cusp suppresses Kozai-Lidov cycles, it does not hinder the precession of the longitude of the ascending node, measured in the reference frame of the gas ring \citep{sub09}. Such precession appears as a change in inclination in the reference frame of the stellar disk. This effect is apparent in Figure \ref{fig:rmsi}, which shows that the root mean square inclination of the stars remains constant if measured from the plane of the inner gas ring.

Moreover, runs A, B and C show higher inclinations at larger radius (Figure \ref{fig:iascatter}). This is due to the fact that the stars on the outer orbits precess faster than those in the inner ones, as expected from the dependence of precession timescale on the semi-major axis ($T_{\rm K} \propto a^{-3/2}$, Equation \ref{eq:tk}). This is in agreement with \citet{map13}.

Since the stars of the disk have different eccentricity and semi-major axis, they precess at different rates, ultimately causing the disk to dismember.
The diffusion of the orbital parameters due to two-body relaxation enhances differential precession, accelerating the dismembering of the disk. To test the importance of two-body relaxation, we run a simulation (named A1) with the same initial conditions as run~A, except that we set the stellar masses to zero. In this way we inhibit two-body relaxation, and the evolution of stellar orbits is driven by the external potentials. 

The inclination distribution in run~A1 at $3 \myr$ is shown in Figure \ref{fig:idistrobackup}. In contrast to run A, the inclinations do not spread and the distribution does not become bimodal. This indicates that the disk changes its orientation without losing coherence and the spiral-shaped tail of normal vectors in Figure \ref{fig:maps} does not form. Thus, two-body relaxation is a key process in understanding the dismembering of a nearly-Keplerian disk. This result is in agreement with the findings of \citet{haa11a}.

{ Since the stellar cusp is modeled as a rigid potential, we neglect additional two-body relaxation between cusp and disk stars. \citet{loc09b} showed that a cusp of stellar remnants enhances the relaxation of angular momentum, increasing orbital eccentricities and disk thickness \citep{loc09}. However, \citet{loc09b} find that that relaxation among disk stars dominates over the relaxation between cusp and disk stars. Thus, we expect that a grainy cusp would enhance the disk disruption in our simulations by a negligible amount.}

\section{Conclusions}\label{sec:conclusions}

We investigate the effect of  gas rings on a nearly-Keplerian stellar disk orbiting a SMBH by means of combined SPH and direct N-body simulations. 
 We simulate the formation of a CNR-like gas ring through the infall and disruption of a molecular gas cloud towards the SMBH. 
In particular, the gas cloud settles down into two concentric rings around the SMBH: the outer ring matches the properties of the CNR in the GC (inner radius $R_{\rm CNR} \sim 1.5 \pc$, mass $M_{\rm CNR} \sim 10^4 \msun$), while the inner ring is less massive ($\sim 10^3 \msun$) and has an outer radius of $\sim 0.4 \pc$.

We make use of the AMUSE software to couple the SPH simulation of the infalling gas cloud to a direct N-body code, which we use to integrate the evolution of the stellar disk.
The stellar disk has properties similar to the CW disk and was formed self consistently by the infall and collapse of a disrupted molecular cloud. 
Our simulations include the effect of the stellar cusp, modeled as a rigid potential.
We simulate different inclinations $\theta_i$ between the infalling gas cloud and the stellar disk: $\theta_i = 10^\circ$ (run A), $45^\circ$ (run B), $90^\circ$ (run C).

We find that the outer ring is inefficient in affecting the stellar orbits on a timescale of $3 \myr$.
On the other hand, the inner ring of gas can significantly affect the stellar disk inclination and coherence.

The inner gas ring induces precession of the longitude of the ascending node $\Omega$ on the disk stars, which appears as a change of the inclinations in the reference frame of the stellar disk. As a consequence, the disk precesses about the axis of symmetry of the inner gas ring. We do not find precession of eccentricity and inclination with respect to the gas ring, because it is suppressed by the stellar cusp.

We find that the precession of $\Omega$ is faster for smaller angles between the stellar disk and the inner gas ring $\theta_{\rm inner}$, as expected from timescale dependence $T_{\rm K} \propto \cos^{-1}(\theta_{\rm inner})$. After $3 \myr$, the stellar disk has changed its orientation by $35^\circ$, $45^\circ$ and $10^\circ$ in runs A ($\theta_{\rm inner} \simeq 20^\circ$), B ($\theta_{\rm inner} \simeq 37^\circ$) and C ($\theta_{\rm inner} \simeq 77^\circ$), respectively. 
Since the inclination changes are driven by $\Omega$-precession, the stellar disk inclination cannot increase more than twice the angle between the gas ring and the stellar disk; as a consequence, the stars in run~B achieve an higher inclination with respect to run~A.

We find that the combined effect of two-body relaxation and $\Omega$-precession can displace stars from the disk. We verified that neither of the two processes can drive the disk dismembering alone. 
Two-body relaxation introduces a spread in the orbital elements of the individual stars of the disk, inducing differential precession. This differential precession leads to the dismembering of the stellar disk, which loses $30\%$ of the stars in run~A ($\theta_{\rm inner} \simeq 20^\circ$) at $3 \myr$. In run~B ($\theta_{\rm inner} \simeq 37^\circ$), the disk lost only $10\%$ of the stars in $3 \myr$, while in run~C ($\theta_{\rm inner} \simeq 77^\circ$) $\Omega$-precession is inefficient and the stellar disk remains coherent. 

In conclusion, our simulations show that the gas in the innermost $0.5 \pc$ (i.e. the inner cavity) can play a crucial role in the evolution of the stellar orbits in the Galactic center.




\acknowledgments
We thank the anonymous referee for their invaluable comments. We thank Alessia Gualandris, Mario Spera, Brunetto Marco Ziosi, Edwin van der Helm, { Bence Kocsis} and Lucie J\'ilkov\'a for useful discussions. 

AAT, MM and AB acknowledge financial support from INAF through grant PRIN-2014-14. MM acknowledges financial support from the Italian Ministry of Education, University and Research (MIUR) through grant FIRB 2012 RBFR12PM1F { and from Fondation MERAC through grant `The physics of gas and protoplanetary discs in the Galactic centre'}. This work was supported by the Interuniversity Attraction Poles Programme initiated by the Belgian Science Policy Office (IAP P7/08CHARM) and by the Netherlands Research Council NWO (grants \#643.200.503, \#639.073.803 and \#614.061.608) and by the Netherlands Research School for Astronomy (NOVA).

\clearpage



\clearpage






\end{document}